\documentclass[3p,times,procedia]{elsarticle}
\usepackage{nupha_ecrc}

\volume{00}

\firstpage{1}

\journalname{Nuclear Physics A}

\runauth{}

\jid{nupha}

\jnltitlelogo{Nuclear Physics A}

\usepackage{amsmath, amsthm, amssymb}
\usepackage{siunitx}

\usepackage{lineno}

\usepackage[figuresright]{rotating}

\begin{document}

\begin{frontmatter}



\dochead{XXVIIth International Conference on Ultrarelativistic Nucleus-Nucleus Collisions\\ (Quark Matter 2018)}

\title{ALICE measurements of flow coefficients and their correlations in small (pp and p--Pb) and large (Xe--Xe and Pb--Pb) collision systems}


\author{Katar\'ina Gajdo\v sov\' a (on behalf of the ALICE Collaboration)}

\address{Niels Bohr Istitute, University of Copenhagen, Blegdamsvej 17, 2100 Copenhagen, Denmark}

\begin{abstract}
Many observables which are used as a signature of the collective effects in heavy-ion collisions when measured in high multiplicity pp and pA interactions reveal a very similar behaviour. We will present first measurements of different order flow coefficients and their magnitude correlations for data collected by ALICE during the LHC Run 2 operation, which includes pp collisions at $\sqrt{s} = \SI{13}{TeV}$, p--Pb at $\sqrt{s_{\rm{NN}}} = \SI{5.02}{TeV}$, Xe--Xe at $\sqrt{s_{\rm{NN}}} = \SI{5.44}{TeV}$ and Pb--Pb collisions at $\sqrt{s_{\rm{NN}}} = \SI{5.02}{TeV}$. Such a broad spectrum of colliding systems with different energies and wide range of multiplicity allow for detailed investigation of their collision dynamics. The measurements are based on a newly developed subevent technique, which was proven to be particularly important for studies in small systems. The results provide an important insight into the nature of collective phenomena in different collision systems.
\end{abstract}

\begin{keyword}
LHC \sep ALICE \sep small systems \sep anisotropic flow \sep cumulants


\end{keyword}

\end{frontmatter}



\section{Introduction}
\label{introduction}

Collisions of heavy ions at ultrarelativistic energies at RHIC and the LHC serve to study the Quark--Gluon Plasma (QGP), a state of QCD matter where quarks and gluons are in a deconfined state. One of the most suitable observables to probe the properties of the QGP are the flow coefficients $v_n = \langle {\rm cos} \, [n(\varphi - \Psi_n)]\rangle$, obtained from the Fourier expansion of the final azimuthal particle distribution with respect to a common symmetry plane $\Psi_n$~\cite{Voloshin:1994mz}:
\begin{equation}
\frac{dN}{d\varphi} \propto 1 + 2 \sum_{n=1}^{\infty} v_ne^{in(\varphi - \Psi_n)}.
\end{equation} 

Observables, believed to indicate the presence of the QGP in heavy ion collisions, revealed similar features in high multiplicity p--Pb and pp collisions, which were originally considered as a reference without the emergence of the QGP. In particular, the measurements of the near side ``ridge'' structure in the di-hadron correlations~\cite{Khachatryan:2010gv} or the negative sign of the four-particle cumulant~\cite{Khachatryan:2015waa} indicate the presence of long-range multi-particle correlations, usually interpreted as collectivity. The origin of these phenomena is still under debate. The results presented here will bring more insight into the nature of collectivity.

\section{Analysis details}
\label{analysisdetails}

The results presented here were obtained from data samples recorded by ALICE~\cite{Aamodt:2008zz} during the LHC Run 2 data taking, in particular from Pb--Pb collisions at $\sqrt{s_{\rm NN}} = 5.02$ TeV, Xe--Xe collisions at $\sqrt{s_{\rm NN}} = 5.44$ TeV, p--Pb collisions at $\sqrt{s_{\rm NN}} = 5.02$ TeV and pp collisions at $\sqrt{s} = 13$ TeV. A minimum-bias trigger requiring a coincidence of signals between the two arrays of the V0 detector (V0A and V0C) was used in all collision systems except for pp collisions, where the data were collected with a dedicated high multiplicity trigger selecting events based on the amplitude in both arrays of the V0 detector. The threshold of the trigger corresponds to events with a multiplicity in the V0 acceptance 4 times larger than the minimum-bias average. Overall, data samples of $310\cdot10^6$ high multiplicity pp collisions, as well as $230\cdot10^6$ p--Pb, $1.3\cdot10^6$ Xe--Xe and $55\cdot10^6$ Pb--Pb minimum-bias collisions were used in the analysis.

Only charged particles with a transverse momentum $0.2 < p_T < 3.0$ GeV/$c$ and with full azimuthal coverage in pseudorapidity $|\eta| < 0.8$ were used for the analysis. Measurements of cumulants~\cite{Borghini:2000sa,Borghini:2001vi,Borghini:2001zr} and Symmetric Cumulants~\cite{Bilandzic:2013kga} were calculated using the generic framework~\cite{Bilandzic:2013kga} with implemented corrections for non-uniform acceptance and tracking inefficiencies by weighting the Q-vectors. The subevent method~\cite{Jia:2017hbm,Huo:2017nms} was employed in the measurements in order to suppress short-range few-particle correlations, denoted as non-flow. 

\section{Results}
\label{results}



\begin{figure}[!htb]
\centering
\begin{minipage}{0.5\textwidth}
\centering
\includegraphics[width=0.9\textwidth]{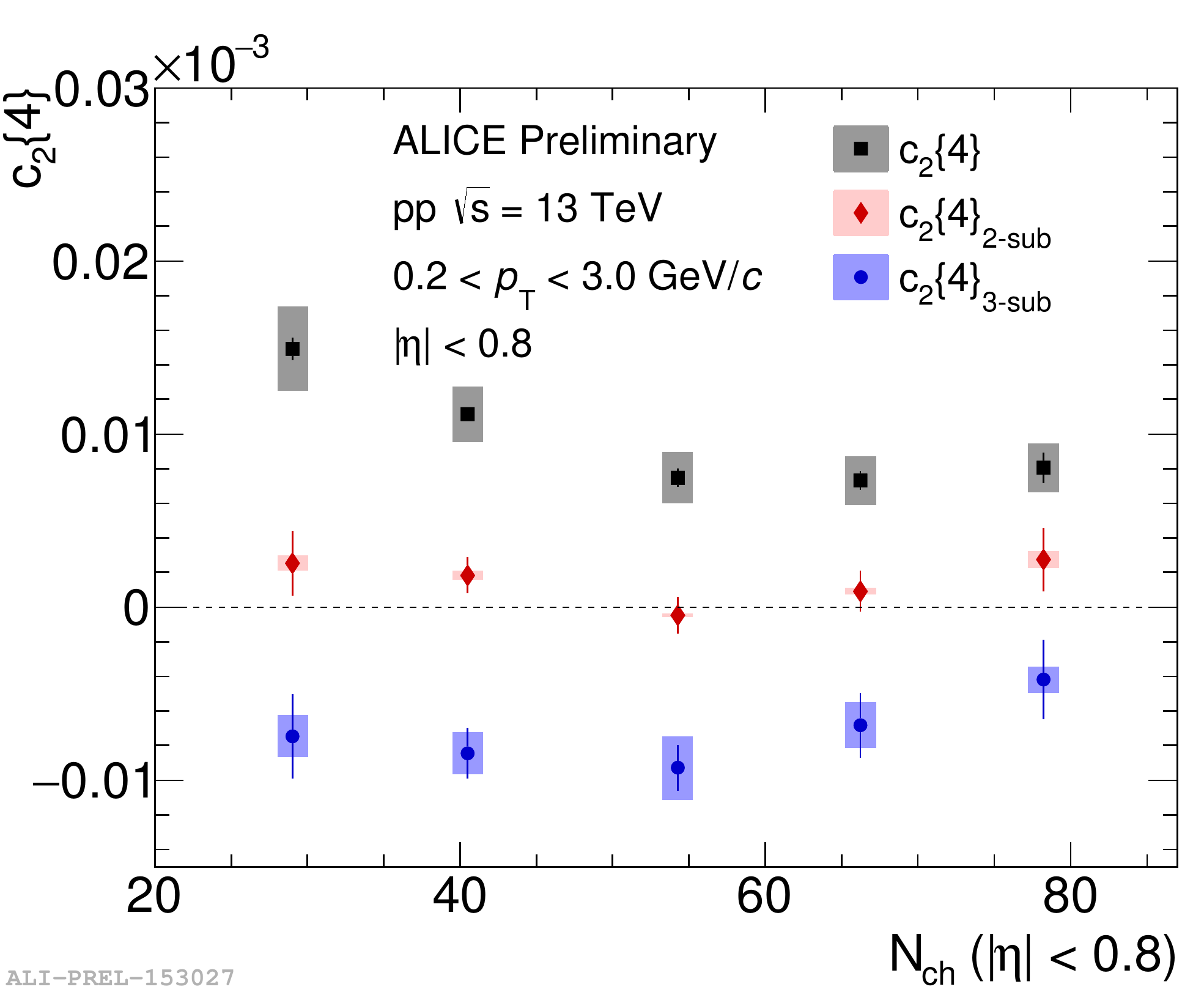}
\end{minipage}\hfill
\begin{minipage}{0.5\textwidth}
\centering
\includegraphics[width=0.9\textwidth]{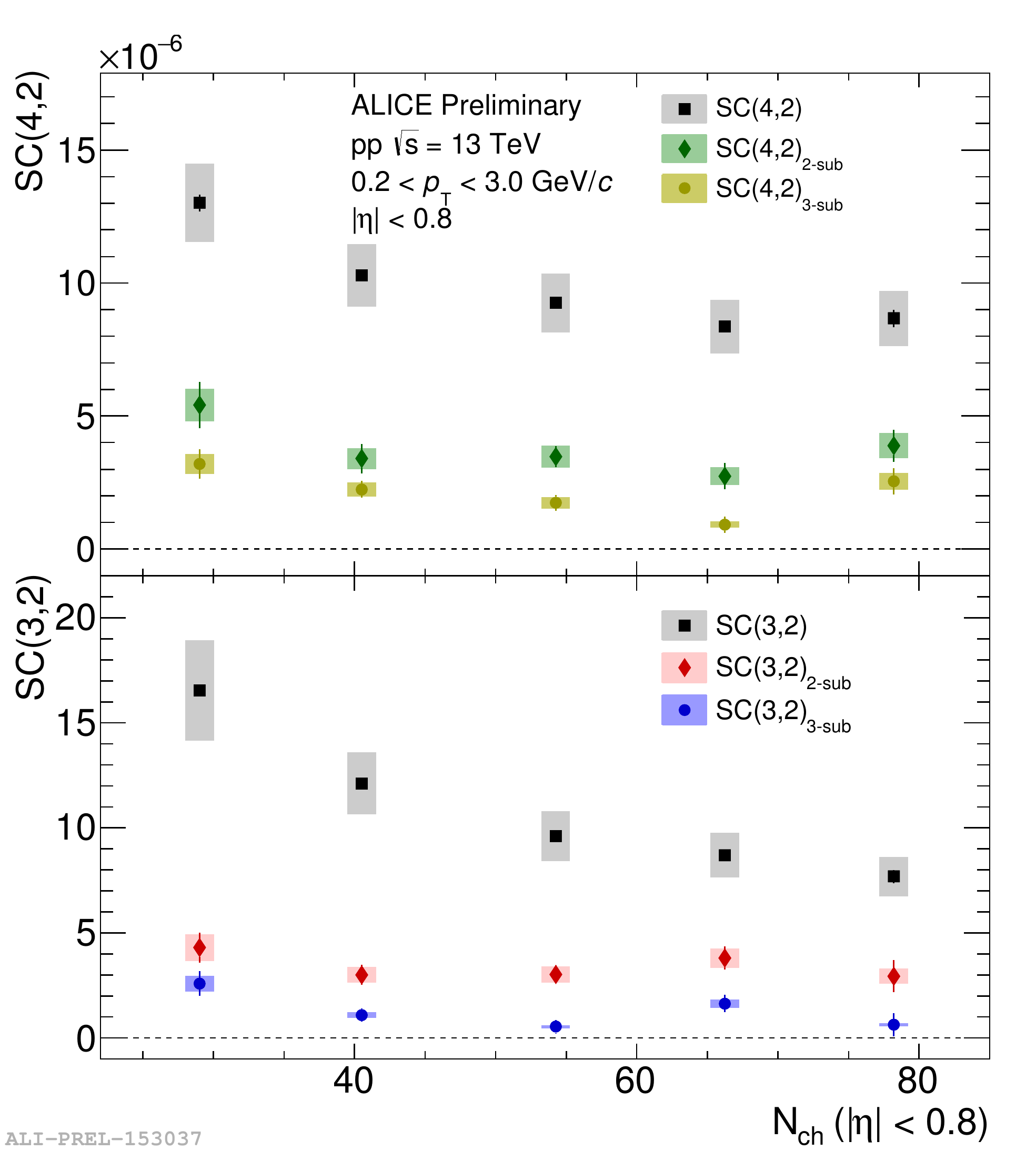}
\end{minipage}
\caption{Left: Multiplicity dependence of $c_2\{4\}$ with the standard, 2-subevent and 3-subevent method in pp collisions. Right: Multiplicity dependence of $SC(4,2)$ and $SC(3,2)$ with the standard, 2-subevent and 3-subevent method in pp collisions.}
\label{fig1}
\end{figure}

Small collision systems, especially pp collisions, are dominated by non-flow correlations. 
While a positive $c_2\{4\}$ is measured in pp collisions,  the $c_2\{4\}_{\rm 3-sub}$ is found to be negative after the suppression of non-flow with the subevent method, shown in Fig.~\ref{fig1} (left). This is measured for the first time in pp collisions with the ALICE detector, and it will allow us to extract a real-valued $v_2\{4\}$. Furthermore, the effects of non-flow in Symmetric Cumulants in pp collisions are investigated in the right panel of Fig.~\ref{fig1}. A strong positive correlation is observed for $SC(4,2)$ and $SC(3,2)$, which significantly decrease with the subevent method. Therefore, it is crucial to report results in small collision systems with the subevent method, which makes them less biased by non-flow correlations.

\begin{figure}[!htb]
\centering
\begin{minipage}{0.5\textwidth}
\centering
\includegraphics[width=1.0\textwidth]{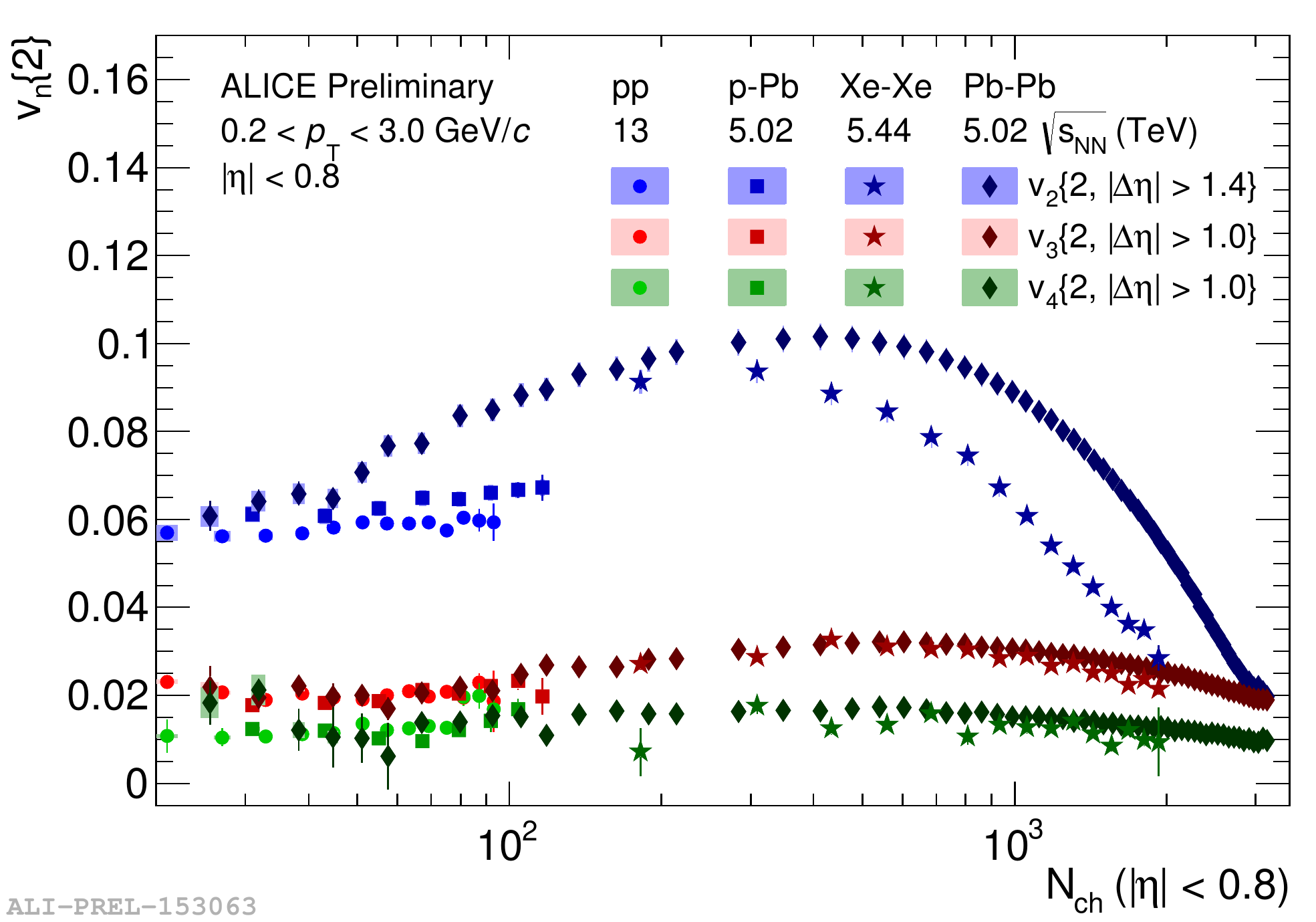}
\end{minipage}\hfill
\begin{minipage}{0.5\textwidth}
\centering
\includegraphics[width=1.0\textwidth]{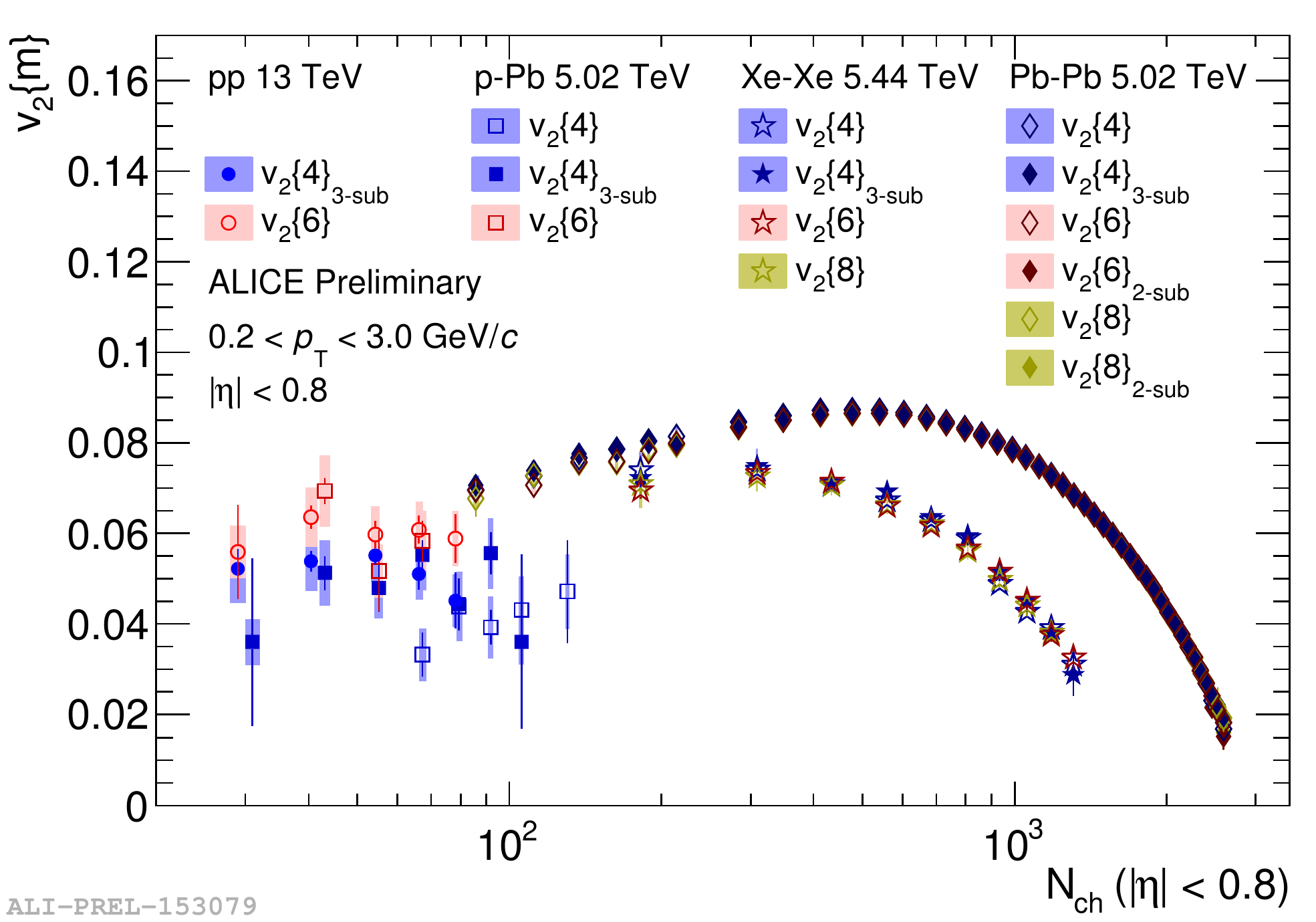}
\end{minipage}
\caption{Left: Multiplicity dependence of $v_2$, $v_3$ and $v_4$ measured using the two-particle cumulant method in small (pp and p--Pb) and large (Xe--Xe and Pb--Pb) collision systems. Right: Multiplicity dependence of $v_2\{m\}$ for $m>2$ in all four collision systems.}
\label{fig2}
\end{figure}

Measurements of flow coefficients $v_n$ using the two particle cumulant from pp, p--Pb, Xe--Xe and Pb--Pb collisions are shown in the left panel of Fig.~\ref{fig2}. An ordering of flow coefficients $v_2 > v_3 > v_4$ is observed in large collision systems, as well as a clear multiplicity dependence of $v_2$ reflecting the initial geometry of the overlapping region of the colliding nuclei. At low multiplicities, the values of $v_n$ from Xe--Xe and Pb--Pb collisions become compatible with those measured in pp and p--Pb collisions, all exhibiting a weak multiplicity dependence. Similarly as in Xe--Xe and Pb--Pb collisions, an ordering of $v_n$ is reported in small collision systems, too. 

The right panel of Fig.~\ref{fig2} shows measurements of $v_2\{m\}$ ($m > 2$) compared for all four collision systems. The measurements in large collision systems are compatible, suggesting the presence of long range ($v_2\{m\} \approx v_2\{m\}_{\rm sub}$) and multi-particle ($v_2\{4\} \approx v_2\{6\} \approx v_2\{8\}$) correlations. Non-flow effects can be further suppressed in multi-particle cumulants in p--Pb collisions resulting in an increase $v_2\{4\}_{\rm 3-sub} > v_2\{4\}$ and extension of this measurement down to lower multiplicities. In pp collisions, a real-valued $v_2\{4\}_{\rm 3-sub}$ is measured for the first time with the ALICE detector. Moreover, measurements of $v_2\{4\}$ in both pp and p--Pb collisions are compatible with $v_2\{6\}$ results, showing the existence of multi-particle correlations in small collision systems. Thus, the observations revealed by the measurements of flow coefficients imply the presence of collectivity in small collision systems. 


\begin{figure}[!htb]
\centering
\includegraphics[width=0.5\textwidth]{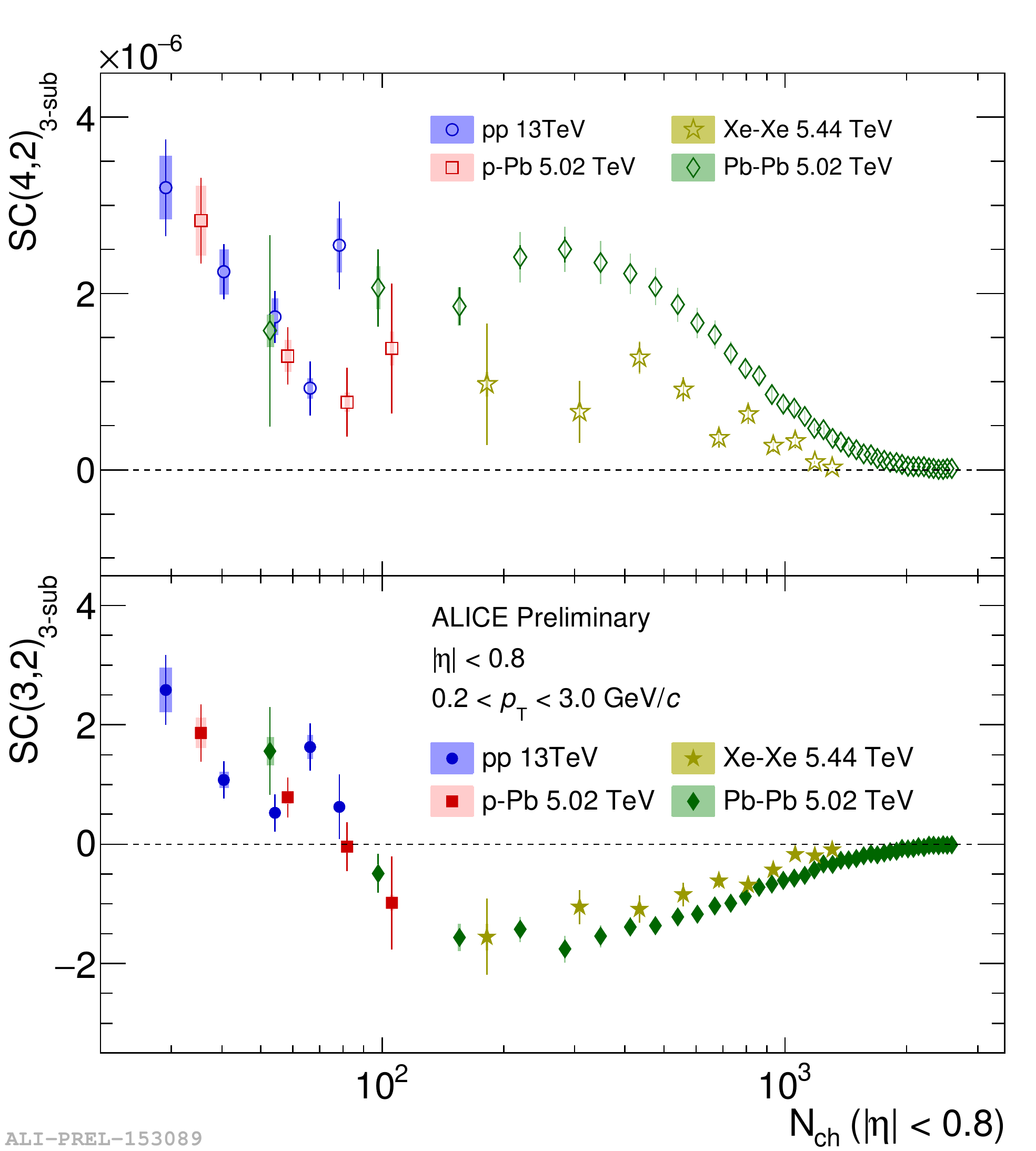}
\caption{Multiplicity dependence of $SC(m,n)_{\rm 3-sub}$ in pp, p--Pb, Xe--Xe and Pb--Pb collisions.}
\label{fig3}
\end{figure}

Various theoretical models, either based on hydrodynamic evolution~\cite{Weller:2017tsr,Mantysaari:2017cni,Zhao:2017rgg}, initial state correlations only~\cite{Dusling:2017dqg}, or other effects~\cite{Bzdak:2014dia,Bierlich:2017vhg}, can reproduce the measurements of the flow coefficients from two-particle correlations in pp or p--Pb collisions at least on a qualitative level. However, more observables which can help to constrain initial conditions and provide the power to disentangle between different approaches are necessary in order to improve our understanding of collectivity in pp and p--Pb collisions. Measurements of $SC(3,2)$ are sensitive to initial conditions while $SC(4,2)$ provides access to to the dynamical evolution of the system~\cite{ALICE:2016kpq}. Since non-flow correlations largely affect these measurements, especially in small collision systems, only the results with the 3-subevent method are presented in Fig.~\ref{fig3}. A positive $SC(4,2)_{\rm 3-sub}$ is observed in the entire multiplicity range for all collision systems, while a negative $SC(3,2)_{\rm 3-sub}$ is seen at large multiplicities in Xe--Xe and Pb--Pb collisions. At low multiplicities, a transition to a positive correlation is observed in Pb--Pb collisions, which seems to be followed by the results obtained in pp and p--Pb collisions.

\section{Summary}
\label{summary}

We present the measurements of flow coefficients and their correlations using multi-particle cumulants. In order to suppress non-flow contamination in these measurements, a subevent method is applied both to the $v_n\{m\}$ and $SC(m,n)$ observables. This results in a negative sign of $c_2\{4\}_{\rm 3-sub}$ in pp collisions, reported for the first time by the ALICE collaboration. Measurements of the flow coefficients in small collision systems reveal similarities with the results in large collision systems, which indicate the presence of collectivity in pp and p--Pb collisions. The nature of the collective phenomena is further studied with the measurements of Symmetric Cumulants. A positive $SC(4,2)_{\rm 3-sub}$ is reported for all collision systems, and a crossing from negative to positive sign of $SC(3,2)_{\rm 3-sub}$ is shown for both small and large collision systems. The results of $v_n\{m\}$ and $SC(m,n)$ presented from four different collision systems, provide a complex set of information for future model comparisons, which could allow us to conclude about the mechanisms responsible for the observed collectivity in small collision systems.



\begin{thebibliography}{00}

\bibitem{Voloshin:1994mz} S. Voloshin, Y.Zhang, Z.Phys. C70 (1996) 665-672.
\bibitem{Khachatryan:2010gv} V. Khachatryan, et al. (CMS Collaboration), JHEP 1009 (2010) 091.
\bibitem{Khachatryan:2015waa} V. Khachatryan, et al. (CMS Collaboration), Phys. Rev. Lett. 115 (1) (2015) 012301.
\bibitem{Aamodt:2008zz} K. Aamodt, et al. (ALICE Collaboration), JINST 3 (2008) S08002.
\bibitem{Borghini:2000sa} N. Borghini, P. M. Dinh, J.-Y. Ollitrault, Phys. Rev. C63 (2001) 054906.
\bibitem{Borghini:2001vi} N. Borghini, P. M. Dinh, J.-Y. Ollitrault, Phys. Rev. C64 (2001) 054901.
\bibitem{Borghini:2001zr} N. Borghini, P. M. Dinh, J.-Y. Ollitrault, 2001. arXiv:nucl-ex/0110016.
\bibitem{Bilandzic:2013kga} A. Bilandzic, C. H. Christensen, K. Gulbrandsen, A. Hansen, Y. Zhou, Phys. Rev. C89 (6) (2014) 064904.
\bibitem{Jia:2017hbm} J. Jia, M. Zhou, A. Trzupek, Phys. Rev. C96 (3) (2017) 034906.
\bibitem{Huo:2017nms} P. Huo, K. Gajdo\v sov\'a, J. Jia, Y. Zhou, Phys. Lett. B777 (2018) 201–206.
\bibitem{Weller:2017tsr} R. D. Weller, P. Romatschke, Phys. Lett. B774 (2017) 351–356.
\bibitem{Mantysaari:2017cni} H. M{\" a}ntysaari, B. Schenke, C. Shen, P. Tribedy, Phys. Lett. B772 (2017) 681–686.
\bibitem{Zhao:2017rgg} W. Zhao, Y. Zhou, H. Xu, W. Deng, H. Song, Phys. Lett. B780 (2018) 495–500.
\bibitem{Dusling:2017dqg} K. Dusling, M. Mace, R. Venugopalan, Phys. Rev. Lett. 120 (4) (2018) 042002.
\bibitem{Bzdak:2014dia} A. Bzdak, G.-L. Ma, Phys. Rev. Lett. 113 (25) (2014) 252301.
\bibitem{Bierlich:2017vhg} C. Bierlich, G. Gustafson, L. Lnnblad, Phys. Lett. B779 (2018) 58–63.
\bibitem{ALICE:2016kpq} J. Adam, et al. (ALICE Collaboration), Phys. Rev. Lett. 117 (2016) 182301.
\end{thebibliography}



\end{document}